\documentclass[secnumarabic,amssymb, nobibnotes, aps, prd, 11pt, abstract]{revtex4-2}
\usepackage{tocloft}
\usepackage[dvips]{graphicx}
\usepackage{amsmath}
\usepackage{subcaption}
\usepackage{hyperref}
\usepackage{filecontents}
\hypersetup{colorlinks,linkcolor={red},citecolor={blue},urlcolor={red}}
\usepackage{setspace}
\usepackage{xcolor}
\pagecolor{white}
\usepackage{float}

\usepackage{multirow}
\color{black}

\begin{document}

\singlespacing

	\title{Phase Transitions and Chaos Bound in Horava Lifshitz Black Holes using Lyapunov Exponents}%

	\author{Mozib Bin Awal$^1$}
	
	\email{$rs_mozibbinawal@dibru.ac.in$}

	\author{Prabwal Phukon$^{1,2}$}
	\email{$prabwal@dibru.ac.in$}
	
	\affiliation{$^1$Department of Physics, Dibrugarh University, Dibrugarh, Assam,786004.\\$^2$Theoretical Physics Division, Centre for Atmospheric Studies, Dibrugarh University, Dibrugarh,Assam,786004.\\}

	\begin{abstract}
We probe the thermodynamic phase structure of four dimensional Horava Lifshitz black holes by Lyapunov exponent analysis. For both massless and massive test particles, the Lyapunov exponent exhibits a multivalued dependence on temperature in regimes with a first-order phase transition, with distinct branches corresponding to small, intermediate, and large black hole phases, and this behaviour disappears at the critical point. The discontinuity in the Lyapunov exponent acts as an effective order parameter with critical exponent $\delta=1/2$, consistent with mean-field universality. We also find that the chaos bound is generically violated below a threshold horizon radius, with the violation occurring within the thermodynamically stable phase and persisting even in the absence of a phase transition. These results establish the robustness and universality of Lyapunov exponents as probes of black hole thermodynamics in alternative theories of gravity.

\end{abstract}
	
	\maketitle
	
\section{Introduction}\label{sec1}
The study of black hole thermodynamics remains one of the most interesting and fascinating field of research in theoretical physics. Its significance was already established during the formative years with the pioneering works of Bekenstein, Hawking, Bardeen and other \cite{Phys,bekens,Hawking,Hawking2,Bardeen}. In the years that followed, further investigations firmly established that black holes exhibit thermodynamic behaviour. This naturally led to the question of whether black holes can also undergo processes analogous to those found in classical thermodynamic systems. One such phenomenon of considerable interest is the occurrence of ``phase transitions'' in black holes. Early studies of this idea were carried out by P.C.W. Davies and P.Hut in \cite{Davies} and \cite{Hut}, respectively. After the discovery of the AdS/CFT \cite{Maldacena} correspondence by Juan Maldacena in $1997$, significant attention was given to Anti de Sitter black holes and their rich thermodynamics. The next ground-breaking discovery in the field was the inclusion of the cosmological constant $\Lambda$ into the first law of black hole thermodynamics by interpreting it as the thermodynamic pressure. This inclusion of the pressure term in the first law started the so called extended phase space thermodynamics of black holes. Following this, quite a number of studies were conducted in this extended phase space formalism. Noteworthy among them was the remarkable similarity of the black hole phase transition to that seen in fluid systems \cite{Kubiz,Hawkpage,Cai,Kastor,Dolan,Dolan2,Dolan3,Kubizna,Xu,Xu2,Zhang}. The conventional approach to investigating phase transitions in black holes has been through the analysis of thermodynamic quantities such as free energy and specific heat. In recent years, however, several alternative methods for studying phase transitions in black holes have gained increasing attention. Among the most well-established and widely explored approaches is the geometric analysis of the thermodynamic state space. In this framework, one of the most commonly used techniques is the study of Ruppeiner geometry \cite{Ruppeiner:2012uc,Miao:2017cyt,Guo:2019oad,Wei:2019yvs,Wang:2019cax,Yerra:2020oph,Yerra:2021hnh}. Another important direction involves examining the topology of the thermodynamic parameter space \cite{Wu:2022whe,Liu:2022aqt,Fan:2022bsq,Gogoi:2023xzy,Ali:2023zww,Saleem:2023oue,Shahzad:2023cis,Chen:2023elp,Bai:2022klw,Yerra:2022alz,Hazarika:2023iwp}, which has proven to be a powerful tool for understanding phase transitions and critical behavior in black hole systems. Beyond these purely theoretical frameworks, several studies have also investigated possible connections between black hole phase transitions and observable astrophysical phenomena. Examples of such approaches include analyses of quasinormal modes (QNMs) \cite{Liu:2014gvf,Zou:2017juz,Zhang:2020khz,Mahapatra:2016dae,Chabab:2016cem}, the behaviour of test particles in circular orbits around black holes \cite{Wei:2017mwc,Wei:2018aqm,Zhang:2019tzi}, and the properties of black hole shadows \cite{Zhang:2019glo,Belhaj:2020nqy}.

In order to study the dynamics of systems that are highly sensitive to the initial conditions, a field known as Chaos Theory was developed by mathematicians and physicists. A key quantity central to chaos theory is the Lyapunov exponent \cite{lyp}, which measures the rate at which nearby trajectories in phase space diverge or converge over time \cite{lyp2}. A positive Lyapunov exponent indicates exponential divergence of trajectories, signifying chaotic behaviour, whereas a negative Lyapunov exponent corresponds to convergence and hence stability within the system. To investigate the phase transitions in physical systems, chaotic dynamics has been found to be quite useful, it has been used to study the phase transitions in both quantum and classical regimes. Noteworthy examples in this regard include the well-known Sachdev-Ye-Kitaev (SYK) model \cite{syk,syk2}, the Dicke model \cite{Dicke}, finite Fermi systems and quantum dots \cite{finite}, as well as models of long-range coupled oscillators \cite{coscll}, among others. 
In recent years, Lyapunov exponents have also been widely used to study chaotic dynamics within the framework of general relativity and black hole physics. A number of works have investigated the motion of particles in static and axisymmetric gravitational backgrounds \cite{static}, as well as in the vicinity of rotating and charged Kerr--Newman black holes \cite{kerr,kerr2}. Chaotic behaviour has also been explored in multi-black hole configurations \cite{multi} and in scenarios where quantum gravity corrections are incorporated into black hole solutions \cite{qg,qg2}. A significant development in the study of quantum chaos was the proposal by Maldacena, Shenker, and Stanford of a universal upper bound on the growth rate of chaos in quantum systems with a large number of degrees of freedom and possessing a semiclassical gravity dual \cite{mss}. This limit, commonly referred to as the MSS bound, is formulated in terms of the Lyapunov exponent $\lambda$. In natural units, the bound is expressed as $\lambda_L \leq 2\pi \tilde{T}$. Subsequent investigations have verified the validity of this bound in various contexts, including the dynamics of massive particles near black hole horizons \cite{hori}, and have highlighted its close relationship with the existence of event horizons \cite{hori2}. Nevertheless, several studies have also reported situations in which violations of the MSS bound may occur \cite{vio,vio2,vio3,vio4}.

The use of Lyapunov exponents to probe the thermodynamic phase structure of black holes was first explored in \cite{first}. In that work, the authors proposed a conjecture establishing a possible relationship between Lyapunov exponents and black hole phase transitions. They showed that the Lyapunov exponent associated with massless and massive particles in circular orbits around black holes exhibits a multivalued structure when plotted as a function of temperature. The different branches of the Lyapunov exponent can be interpreted as corresponding to distinct thermodynamic phases of the black hole. It was further observed that this multivalued behaviour disappears at a particular critical value of certain model-dependent parameters. Moreover, the discontinuity in the Lyapunov exponent across the transition was shown to serve as an order parameter, characterized by a critical exponent of $1/2$ in the vicinity of the phase transition point. Following this initial work, several subsequent studies have verified and expanded upon these findings \cite{le,le2,le3,le4,le5,le6,le7,le8,Awal,le9,le10,mba}.

In this work we, extend the analysis of the interconnection of Lyapunov exponent and phase transitions in black holes beyond general relativity by considering a particular type of black holes in $D=4$ Horava Lifshitz gravity. Horava gravity is widely regarded as a potential candidate for a quantum theory of gravity at very high energy scales \cite{Horava:2009uw}. Consequently, Horava-Lifshitz (HL) black hole solutions have attracted significant attention, particularly in studies of their thermodynamics and phase transitions \cite{ht1,ht2,ht3,ht4}, owing to their rich and intricate phase structures. The organisation of the manuscript is done in the following ways. In section \ref{sec2}, we elaborate the methodology to calculate the Lyapunov exponents revolving in unstable circular orbits. In section \ref{sec3} we analyze the thermodynamics of the $D=4$ Horava Lifshitz black hole. In section \ref{sec4} we demonstrated how the Lyapunov exponent associated with massive and massless particles probe the thermodynamic phase transition of $D=4$ Horava Lifshitz black hole. In section \ref{sec5} we calculate the discontinuity in the Lyapunov exponent and show that the critical exponent associated with it is $1/2$. In section \ref{sec6} we investigate the chaos bound and its violation. Our results are summarised in section \ref{sec7}

\section{Review of Lyapunov exponents}\label{sec2}
In this section, we provide a brief overview of the procedure used to compute the Lyapunov exponent. We consider the dynamics of both massless and massive particles undergoing unstable circular motion around the black hole. To begin with, we write down the Lagrangian that governs the geodesic motion of such particles in the given spacetime background, restricting the motion to the equatorial plane $\theta=\pi/2$.
\begin{equation}\label{eq1}
2\mathcal{L}=-f(r)\dot{t}^2+\frac{\dot{r}^2}{f(r)}+r^2\dot{\phi^2}
\end{equation}
with the dot denoting differentiation with respect to the proper time $\tau$. Using the standard relation $p_q=\frac{\partial\mathcal{L}}{\partial\dot{q}}$, we derive the canonical momenta corresponding to the generalized coordinates, which gives,
\begin{equation}\label{eq2}
p_t=\frac{\partial\mathcal{L}}{\partial\dot{t}}=-f(r)\dot{t}=-E,\; p_r=\frac{\partial\mathcal{L}}{\partial\dot{r}}=\frac{\dot{r}}{f(r)},\; p_\phi=\frac{\partial\mathcal{L}}{\partial\dot{\phi}}=r^2\dot{\phi}=L
\end{equation}
with $E$ and $L$ being the conserved energy and angular momentum of the particle. Equation (\ref{eq2}) gives \begin{equation}\label{eq3}
\dot{t}=\frac{E}{f(r)} \quad \dot{\phi}=\frac{L}{r^2}
\end{equation}
We then use the standard definition of Hamiltonian and the relation given in (\ref{eq3}) to write,
\begin{equation}\label{eq4}
2\mathcal{H}=-E\dot{t}+\frac{\dot{r}^2}{f(r)}+\frac{L^2}{r^2}=\delta_1
\end{equation}
In the above equation, $\delta_1 = -1$ and and $\delta_1 = 0$ corresponds to timelike and null geodesics respectively. For a particle in radial motion, the effective potential is given by the equation,
\begin{equation}\label{eq5}
V_{\text{eff}}=f(r)\left[\frac{L^2}{r^2}+\frac{E^2}{f(r)}-\delta_1\right]
\end{equation}
The angular momentum can be written in terms of the effective potential by setting $E=0$ in Eq.~(\ref{eq5}). Substituting this result into Eq.~(\ref{eq4}) allows the Hamiltonian to be expressed as
\begin{equation}\label{eq6}
\mathcal{H}=\frac{V_{\text{eff}}-E^2}{2f(r)}+\frac{f(r)p^2_r}{2}+\frac{\delta_1}{2}
\end{equation}
Therefore, the equation of motion can be expressed as follows in the proper time configuration 
\begin{equation}\label{eq7}
\dot{r}=\frac{\partial\mathcal{H}}{\partial p_r}=f(r)p_r,\quad \dot{p_r}=\frac{\partial\mathcal{H}}{\partial r}=-\frac{V'_{\text{eff}(r)}}{2f(r)}-\frac{f'(r)p^2_r}{2}+\frac{V_{\text{eff}}-E^2}{2f^2(r)}f'(r)
\end{equation}
Here, the prime symbol denotes differentiation with respect to the radial coordinate $r$. Expanding the equations of motion to first order about the circular orbit located at $r_0$, one can construct the associated stability matrix $K$. This matrix is defined with respect to the coordinate time $t$ and takes the form,
\begin{equation}\label{eq8}
K = \begin{pmatrix}
0 & \frac{f(r_0)}{\dot{t}} \\
-\frac{V''_{\text{eff}}(r_0)}{2f(r_0)\dot{t}} & 0
\end{pmatrix}
\end{equation}
The eigenvalue of the stability matrix above, gives us the Lyapunov exponent
\begin{equation}\label{eq9}
\lambda=\sqrt{-\frac{V''_{\text{eff}}(r_0)}{2\dot{t}^2}}
\end{equation}
here, for simplicity, we may drop the $\pm$ sign before the square root.
The criterion determining the instability of circular geodesics can be written as
\begin{equation}\label{eq10}
V'_{\text{eff}}(r_0)=0,\quad V''_{\text{eff}}(r_0)<0
\end{equation}
which simultaneously determines the radius of the unstable circular orbit, $r_0$. Substituting these two relations into Eq.~(\ref{eq5}), we obtain \begin{equation}\label{eq11}
\frac{E}{L}=\frac{\sqrt{f(r_0)}}{r_0}
\end{equation}
Substituting this expression into Eq.~(\ref{eq3}) and setting $\delta_1 = 0$ (which corresponds to massless particles), we obtain \begin{equation}\label{eq12}
\dot{t}=\frac{L}{r_0\sqrt{f(r_0)}}
\end{equation}
This allows us to arrive at the final expression for the Lyapunov exponent, given in Eq.~(\ref{eq9}), as
\begin{equation}\label{eq13}
\lambda=\sqrt{-\frac{r^2_0f(r_0)}{2L^2}V''_{\text{eff}}(r_0)}
\end{equation}
Adopting the same techniques, for massive particles ($\delta_1 = 1$), the corresponding relations can be obtained as
\begin{equation}\label{eq14}
E^2=\frac{2f^2(r_0)}{2f(r_0)-r_0f'(r_0)}\quad \text{and} \quad L^2=\frac{r^3_{0}f'(r_0)}{2f(r_0)-r_0f'(r_0)}
\end{equation}
Therefore, from equation (\ref{eq3}) we have
\begin{equation}\label{eq15}
\dot{t}=\frac{1}{\sqrt{f(r_0)-\frac{1}{2}r_0f'(r_0)}}
\end{equation}
Finally, by making use of Eq.~(\ref{eq9}), we obtain the corresponding expression for the Lyapunov exponent in the case of massive particles.
\begin{equation}\label{eq16}
\lambda=\frac{1}{2}\sqrt{\left[r_0f'(r_0)-2f(r_0)\right]V''_{\text{eff}}}
\end{equation}

\section{horava lifshitz black hole}\label{sec3}
\begin{figure}[h!]
        \includegraphics[width=0.5\textwidth]{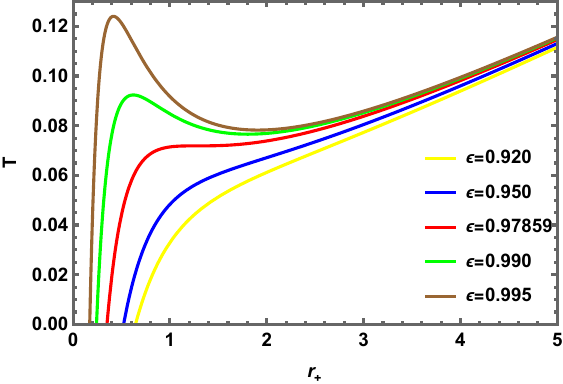}
        \caption{Hawking temperature as a function of the horizon radius for different values of $\epsilon$}
        \label{f1}
\end{figure}

\begin{figure}[h!]
    \centering
    \begin{subfigure}[b]{0.45\textwidth}
        \centering
        \includegraphics[width=\textwidth]{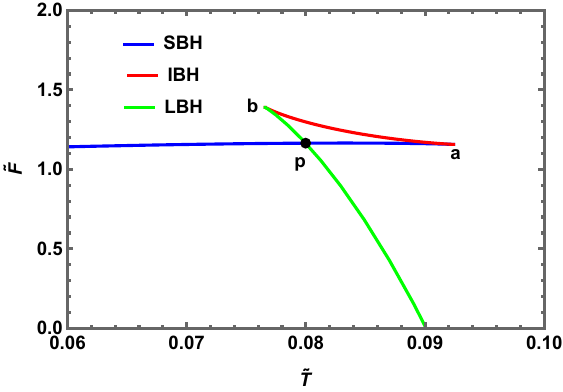}
        \caption{$\epsilon=0.99>\epsilon_c$}
        \label{f2a}
    \end{subfigure}
    
    \vskip\baselineskip  

    \begin{subfigure}[b]{0.45\textwidth}
        \centering
        \includegraphics[width=\textwidth]{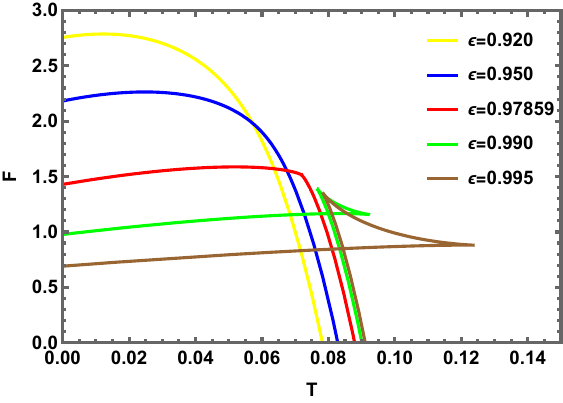}
        \caption{free energy with different values of $\epsilon$}
        \label{f2b}
    \end{subfigure}
    \hspace{0.8cm}
    \begin{subfigure}[b]{0.45\textwidth}
        \centering
        \includegraphics[width=\textwidth]{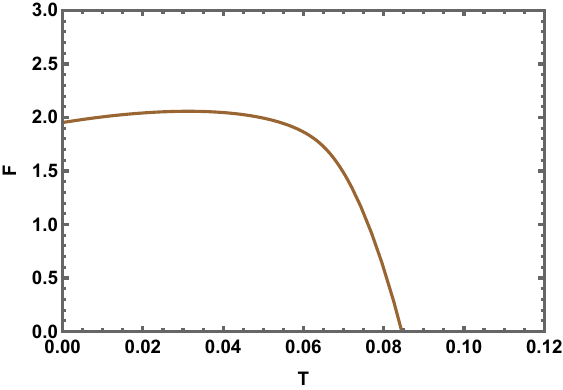}
        \caption{$\epsilon<\epsilon_c$}
        \label{f2c}
    \end{subfigure}

    \caption{Gibbs free energy as a function of temperature}
    \label{f2}
\end{figure}

In this section we introduce the Horava Lifshitz black hole and derive the necessary thermodynamic quantities. We start by writing the space-time metric for the $D=d+1$ dimensional AdS Horava Lifshitz black hole \citep{ht1,Li}
\begin{equation}\label{eq17}
ds^2=-f(r)dt^2+\frac{dr^2}{f(r)}+r^2 d\Omega^2_{d-1,k}
\end{equation}
where the metric function $f(r)$ is given by 
\begin{equation}\label{eq18}
f(r)=k+\frac{32 \pi  P r^2}{(d-1) d \left(1-\epsilon ^2\right)}-4 r^{2-\frac{d}{2}} \sqrt{\frac{64 \pi ^2 P^2 \epsilon ^2 r^d}{(d-1)^2 d^2 \left(1-\epsilon ^2\right)^2}+\frac{\pi  (d-2) M P}{d \left(1-\epsilon ^2\right)}}
\end{equation}
In the above lapse function, $M$ and $P$ are the mass and thermodynamic pressure of the black hole, $d$ is the spatial dimension, $\epsilon$ is a constant appearing in the action of the Horava Lifshitz gravity \citep{Li} and $k$ can take the values $+1, 0, -1$ corresponding to spherical, flat and hyperbolic horizons respectively. In what follows, we shall consider the spatial dimension to be $3$ ($d=3$) and $k=1$ - the spherical horizon case. Imposing those, and setting $f(r_+)=0$, we get the mass of the black hole, 
\begin{equation}\label{eq19}
M=\frac{\left(16 \pi  P r_+^2+3\right)^2-9 \epsilon ^2}{48 \pi  P r_+}
\end{equation}
The Hawking temperature is given by,
\begin{equation}\label{eq20}
T=\frac{256 \pi ^2 P^2 r_+^4+32 \pi  P r_+^2+3 \epsilon ^2-3}{8 \pi  r_+ \left(16 \pi  P r_+^2-3 \epsilon ^2+3\right)}
\end{equation}
The entropy for the Horava Lifshitz black hole is calculated as,
\begin{equation}\label{eq21}
S=\begin{cases}
			4 \pi  r_{+}^2 \left(1+\frac{d k \left(1-\epsilon ^2\right) \ln (r_{+})}{8 \pi  P r_{+}^2}\right)+S_{0}& \text{d=3}\\
			\frac{16 \pi  r_{+}^{d-1} \left(\frac{d (d-1)^2 k \left(1-\epsilon ^2\right)}{32 \pi  (d-3) P r_{+}^2}+1\right)}{(d-2) (d-1)^2}+S_{0}& \text{$d\geq 3$}
			
		\end{cases}
\end{equation}
Here, $S_0$ is an integration constant and we may fix it to $0$. For this particular study, we have the following entropy expression
\begin{equation}\label{eq22}
S=4 \pi  r_+^2 \left(\frac{3 \left(1-\epsilon ^2\right) \log (r_+)}{8 \pi  P r_+^2}+1\right)
\end{equation}
We can calculate the Gibbs free energy using the definition $F=M-TS$ giving us
\begin{equation}\label{eq23}
F=\frac{\left(16 \pi  P r_+^2+3\right){}^2-9 \epsilon ^2}{48 \pi  P r_+}-\frac{r_+ \left(32 \pi  P r_+^2 \left(8 \pi  P r_+^2+1\right)+3 \left(\epsilon ^2-1\right)\right) \left(1-\frac{3 \left(\epsilon ^2-1\right) \log \left(r_+\right)}{8 \pi  P r_+^2}\right)}{2 \left(16 \pi  P r_+^2-3 \epsilon ^2+3\right)}
\end{equation}
To find the critical points of the different thermodynamic quantities, we have to solve the following set of equations simultaneously
\begin{equation}\label{eq24}
\frac{\partial T}{\partial r_+}=\frac{\partial^2T}{\partial r_+^2}=0
\end{equation}
Using the Hawking temperature in equation \ref{eq20}, and solving the above equations, we get the following critical values,
\begin{equation}\label{eq25}
r_c=1.2783028,\quad \epsilon_c=0.9785932,\quad T_c=0.0718829
\end{equation}
The variation of the Hawking temperature (T) with horizon radius ($r_+$) is shown in figure \ref{f1}
In figure \ref{f1}, the brown and the green curve corresponds to the $\epsilon>\epsilon_c$, whereas, the blue and the yellow curve corresponds to $\epsilon<\epsilon_c$. The red curve represents the temperature profile for the critical value of $\epsilon$. It can be observed from here that for $\epsilon$ values less than the critical values, there are two turning points in the temperature profile and correspondingly three black hole branches which are namely - Small Black Hole (SBH), Large Black Hole (LBH) and Intermediate Black Hole (IBH) branch. These three branches can be distinctly shown using the free energy profile of the black hole which is shown in figure \ref{f2a}. Here we can observe the three branches shown in distinct colours - displaying a swallow-tail behaviour which is a characteristic of the first order phase transition. The three branches of the black hole are observed to coexist for $T_b<T<T_a$, where $T_a$ and $T_b$ are the temperature of the extreme points of the free energy curve. At temperature $T_p$, at point $p$, we have a first order phase transition between the small and the large black hole phases. Throughout this range, the free energy of the intermediate black hole branch remains higher than that of both the small and large black hole phases. This renders the IBH state thermodynamically unfavourable and signals its instability within the overall phase structure.
Figure \ref{f2b} and \ref{f2c} shows that the swallow-tail persists for values $\epsilon>\epsilon_c$ and it disappears as soon as $\epsilon<\epsilon_c$ and in that case, we have a free energy curve that is smooth and continuous.

\section{Lyapunov exponents and phase transitions of Horava Lifshitz black holes}\label{sec4}
In this section we compute the Lyapunov exponents associated with both massless and massive particles in unstable circular orbits around the Horava Lifshitz black hole. For massless particles, we do everything analytically, but for the massive particle case, we have to take a numerical approach for the computation of the Lyapunov exponent. We also demonstrate how the thermal profile of the Lyapunov exponent can act as a probe for the black hole's phase transition. We start the analysis using the massless particle case.

\subsection{Masasless Particles}
\begin{figure}[h!]
    \centering
    \begin{subfigure}[b]{0.45\textwidth}
        \includegraphics[width=\textwidth]{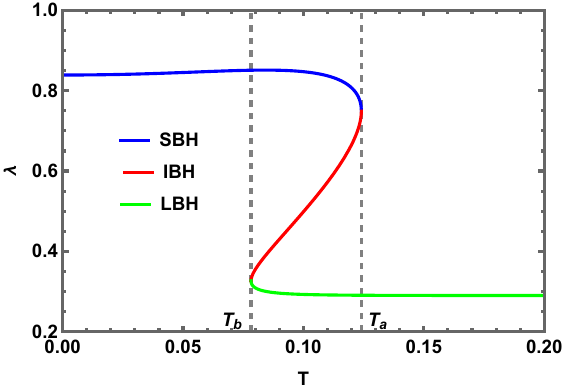}
        \caption{$\epsilon=0.995>\epsilon_c$}
        \label{f3a}
    \end{subfigure}
    \hspace{0.05\textwidth}
    \begin{subfigure}[b]{0.45\textwidth}
        \includegraphics[width=\textwidth]{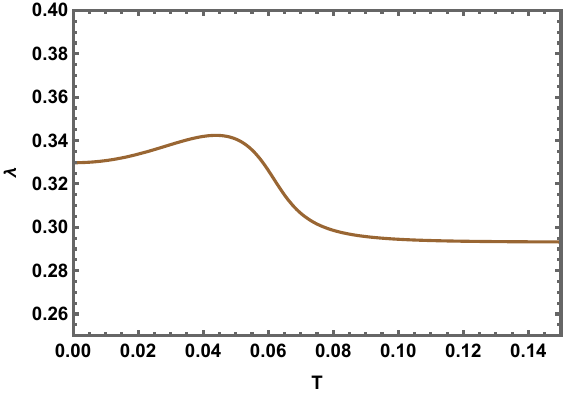}
        \caption{$\epsilon<\epsilon_c$}
        \label{f3b}
    \end{subfigure}
    
    \vspace{0.5cm} 

    \begin{subfigure}[b]{0.45\textwidth}
        \includegraphics[width=\textwidth]{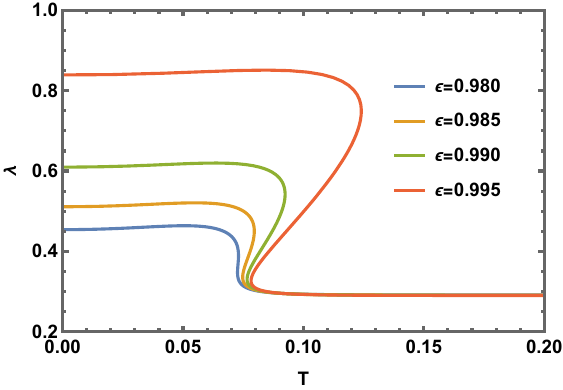}
        \caption{Lyapunov exponent versus temperature}
        \label{f3c}
    \end{subfigure}
    \hspace{0.05\textwidth}
    \begin{subfigure}[b]{0.45\textwidth}
        \includegraphics[width=\textwidth]{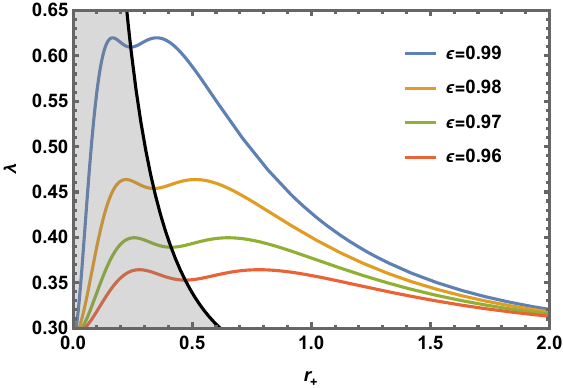}
        \caption{Lyapunov exponent versus horizon radius}
        \label{f3d}
    \end{subfigure}
    \caption{Lyapunov exponent $\lambda$ of massless particles as a function of temperature and horizon radius}
    \label{f3}
\end{figure}

The Lyapunov exponent for massless particle in unstable circular orbit around a black hole can be calculated using equation (\ref{eq5}) and (\ref{eq13}). The analytical expression is hardy any intuitive and extremely cumbersome, so, we directly provide the plot of the Lyapunov exponent as a function of Hawking temperature and horizon radius in figure \ref{f3}. The thermal profile of the Lyapunov exponent is shown in figure \ref{f3a}, where we have used the same colour coding as used in figure \ref{f2a} distinguishing small, large and intermediate black holes. 

We can observe that the Lyapunov exponent shows a distinctive multivaluedness in the temperature range $T_b$ to $T_a$ when the value of $\epsilon$ is greater than the critical value. Within this temperature range the blue and the green curves correspond to small and large black hole branch respectively and the red part of $\lambda$ represents the intermediate black hole branch. The behaviour of the Lyapunov exponent exhibits a strong resemblance to that of the Gibbs free energy, suggesting its usefulness as a diagnostic tool for detecting phase transitions in black holes. Earlier in figure \ref{f2c} we observed that the swallow-tail disappeared for $\epsilon<\epsilon_c$, correspondingly in the Lyapunov exponent profile, we observe in figure \ref{f3b} that the multivalued nature disappears for $\epsilon<\epsilon_c$. The temperature dependence of $\lambda$ for various values of $\epsilon$ is shown in figure \ref{f3c}, where, evidently we also observe the same results. To gain deeper insight into the behaviour of the Lyapunov exponent, we present its variation with the horizon radius $r_+$ in figure \ref{f3d}. The grey-shaded region represents the non-physical region, where the Hawking temperature becomes negative, and the black curve corresponds to the case $\tilde{T} = 0$.

\subsection{Massive Particles}
We now begin our analysis of the Lyapunov exponent corresponding to the massive particles. In this case we have to use equations (\ref{eq5}) and (\ref{eq16}). For massive test particles, the Lyapunov exponents depends on $r_0$ as well as the angular momentum $L$ (we take $L=20$ as a representative choice). However, for the massive particles case, it is not possible to solve the equations (\ref{eq10}) - which are used to determine the radius of the unstable circular orbit $r_0$. 
\begin{figure}[t]
    \centering
    \begin{subfigure}[b]{0.45\textwidth}
        \includegraphics[width=\textwidth]{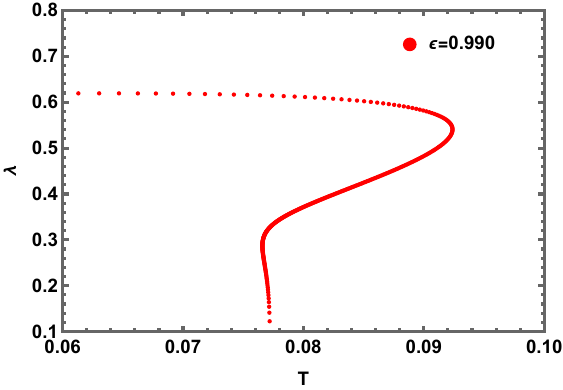}
        \caption{$\epsilon=0.995>\epsilon_c$}
        \label{f4a}
    \end{subfigure}
    \hspace{0.05\textwidth}
    \begin{subfigure}[b]{0.45\textwidth}
        \includegraphics[width=\textwidth]{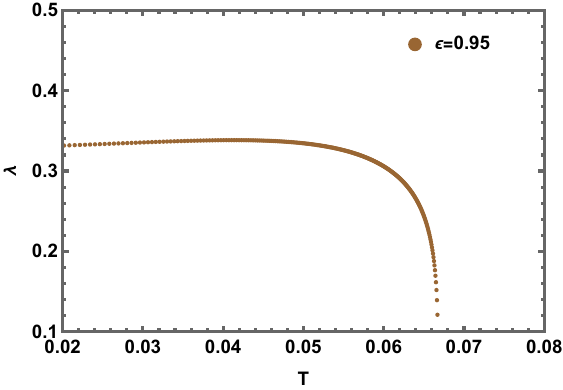}
        \caption{$\epsilon<\epsilon_c$}
        \label{f4b}
    \end{subfigure}
    \caption{Lyapunov exponent $\lambda$ of massive particles as a function of temperature}
    \label{f4}
\end{figure}
Therefore, it is not possible to obtain an analytical expression of the Lyapunov exponent. So, we resort to numerical techniques to compute $\lambda$. We plot the numerical results in figure \ref{f4}. In the left panel, the temperature dependence of the Lyapunov exponent is illustrated for $\epsilon > \epsilon_c$. Similar to the massless case, the Lyapunov exponent exhibits a multivalued structure for $\epsilon>\epsilon_c$, which persists within the temperature range bounded by the two turning points of the curve. And as $\epsilon$ is decreased below $\epsilon_c$, the multivalued nature vanishes, leaving the profile of $\lambda$ only a continuous, smooth curve. A distinctive behaviour of the massive particle case is that in this case the Lyapunov exponent approaches to zero in the large black hole phase. This behaviour indicates that, in the large black hole regime, the motion of massive particles becomes effectively non-chaotic, with $\lambda$ approaching zero, thereby reflecting the absence of unstable equilibrium points. The scenario is persistent even when the value of $\epsilon$ is smaller than $\epsilon_c$ as is illustrated in figure \ref{f4b}. Since an explicit expression for the radius of the unstable circular orbit cannot be obtained, it is not possible to express $\lambda$ analytically in terms of the horizon radius $r_+$. Consequently, its variation with $r_+$ cannot be analyzed in the same manner as in the massless particle case.

\section{Critical exponents for Horava Lifshitz black holes with lyapunov exponent}\label{sec5}
Within the framework of Landau’s theory, an order parameter is a physical quantity that differentiates between two phases, typically vanishing in one phase while acquiring a non-zero value in the other. Familiar examples include magnetization in ferromagnetic systems and the density contrast in liquid-gas transitions. In the context of small-large black hole phase transitions, it has been observed that an analogous role can be played by variations in the Lyapunov exponent, thereby functioning as an effective order parameter. At the phase transition temperature $T_p$, the Lyapunov exponent associated with the small black hole branch is denoted by $\lambda_s$, while that corresponding to the large black hole branch is $\lambda_l$. At the critical point, where $T_p = T_c$, these quantities coincide, i.e., $\lambda_s = \lambda_l = \lambda_c$, where $\lambda_c$ is the value of the Lyapunov exponent evaluated at the critical thermodynamic parameters. 

The difference $\Delta \lambda = \lambda_s - \lambda_l$ then serves as the order parameter, which approaches zero as the system reaches the critical point. To characterize the critical behaviour of $\lambda$, one can evaluate the associated critical exponent, which quantifies how a physical quantity scales in the vicinity of the phase transition. In this work, we adopt the method outlined in \cite{le2,ce} to compute the exponent $\delta$, which satisfies the following equation,
\begin{equation}\label{eq26}
\Delta\lambda\equiv\lambda_s-\lambda_l=\mid\tilde{T}-\tilde{T}_c\mid^\delta
\end{equation}
Now, we express the horizon radius and the Hawking temperature at the phase transition point,
 \begin{equation}\label{eq27}
r_p=r_c\left(1+\Delta\right) \quad \text{and} \quad T(r_+)=T_c\left(1+\epsilon\right)
\end{equation}
where $r_p$ is the horizon radius at the phase transition point, $\mid\Delta\mid\ll1$ and $\mid\epsilon\mid\ll1$. We then expand the Lyapunov exponent in a Taylor series around the critical point $r_c$ to get,
\begin{equation}\label{eq28}
\lambda=\lambda_c+\left[\frac{\partial \lambda}{\partial r_+}\right]dr_++\mathcal{O}\left(r_+\right)
\end{equation}
where, the subscript $c$ is used to denote quantities evaluated at the critical point. So, using the equations (\ref{eq27}) and (\ref{eq28}), we obtain the following equation,
\begin{equation}\label{eq29}
\frac{\Delta\lambda}{\lambda_c}=\frac{\lambda_s-\lambda_l}{\lambda_c}=\frac{r_c}{\lambda}_c\left[\frac{\partial\lambda}{\partial r_+}\right]_c\left(\Delta_s-\Delta_l\right)
\end{equation}
where we have also used the condition that at the critical point, $\lambda_s = \lambda_l = \lambda_c$ and hence $\lambda_s(r_c)-\lambda_l(r_c)=0$.
In a similar manner, we expand the Hawking temperature in a Taylor series about the critical horizon radius $\tilde{r}_c$, which yields
\begin{equation}\label{eq30}
T=T_c+\frac{r^2_c}{2}\left[\frac{\partial^2 T}{\partial r^2_+}\right]\Delta^2
\end{equation}
In this approximation, higher-order terms are neglected, and the condition $\left[\frac{\partial \tilde{T}}{\partial \tilde{r}_+}\right]_c \to 0$ is imposed. Employing equations (\ref{eq29}) and (\ref{eq30}), we subsequently obtain a simplified expression,
\begin{equation}\label{eq31}
\frac{\Delta\lambda}{\lambda_c}=k\sqrt{t-1}
\end{equation}
where $t=\frac{T}{T_c}$. Also,
\begin{equation}\label{eq32}
k=\frac{\sqrt{T_c}}{\lambda_c}\left[\frac{\partial\Delta\lambda}{\partial r_+}\right]_c\left[\frac{1}{2}\frac{\partial^2T}{\partial r^2_+}\right]^{-1/2}_c
\end{equation}
It therefore follows that, in the vicinity of the critical point, the critical exponent $\delta$ corresponding to the order parameter $\Delta \lambda$ assumes the value $1/2$.

\subsection{Numerical Verification}
We now present a numerical verification of the above result in the context of Horava Lifshitz black holes, for both massless and massive particle scenarios. To this end, we plot the scaled order parameter, $\Delta \lambda / \lambda_c$, as a function of the reduced transition temperature $t =T_p / T_c$. 
\begin{figure}[h!]
    \centering
    \begin{subfigure}[b]{0.45\textwidth}
        \includegraphics[width=\textwidth]{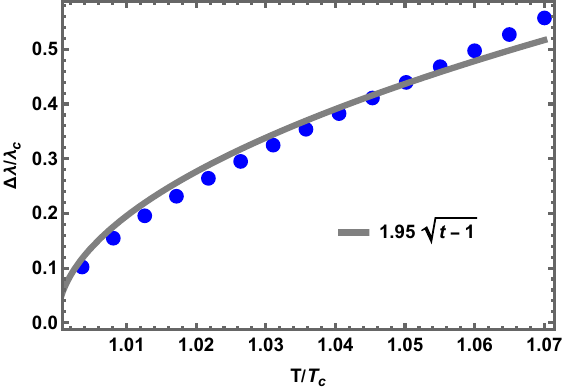}
        \caption{for massless particle}
        \label{f5a}
    \end{subfigure}
    \hspace{0.05\textwidth}
    \begin{subfigure}[b]{0.45\textwidth}
        \includegraphics[width=\textwidth]{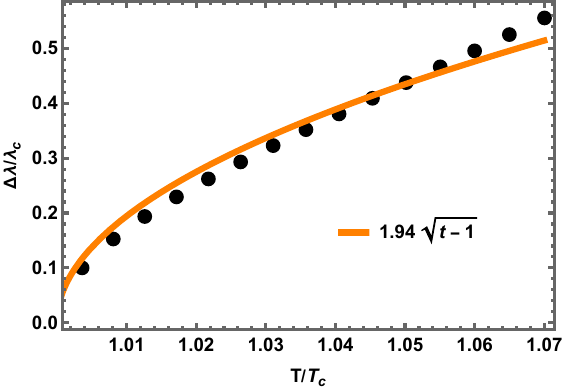}
        \caption{for massive particle}
        \label{f5b}
    \end{subfigure}
    \caption{Rescaled discontinuity in Lyapunov exponent, $\Delta\lambda/\lambda_c$ versus rescaled phase transition temperature $t$}
    \label{f5}
\end{figure}

For the massless case, figure \ref{f5a} illustrates the variation of $\Delta \lambda / \lambda_c$ with $t$ in the vicinity of the critical point. The discrete markers represent numerical data, while the smooth coloured curves correspond to fitted profiles. An analogous analysis is carried out for massive particles, with the corresponding behaviour of $\Delta \lambda / \lambda_c$ as a function of $T_p / T_c$ shown in figure \ref{f5b}. 

Our results indicate that the relation $\Delta \lambda / \lambda_c = k \sqrt{t - 1}$ provides an good enough fit to the numerical data in both cases, with $k = 1.95345$ for massless particles and $k = 1.94348$ for massive particles. Although the proportionality constant $k$ differs between the two cases, the critical exponent $\delta$ consistently retains the value $1/2$.

\section{Chaos Bound and its violation}\label{sec6}
In the context of black hole physics, the Lyapunov exponent is constrained by an upper limit, widely known as the chaos bound \cite{mss}. This bound, expressed as $\lambda \le \frac{2\pi T}{\hbar}$, sets the maximum rate at which chaotic behaviour can develop in thermal quantum systems. It was originally derived by Maldacena, Shenker, and Stanford using arguments rooted in quantum field theory, along with considerations involving shock wave dynamics near event horizons. 

Subsequently, Hashimoto and Tanahashi reformulated this bound within the framework of particle dynamics by examining the motion of test particles in the near-horizon region. Their analysis led to the condition $\lambda \le \kappa$, where $\kappa$ represents the surface gravity \cite{Hashimoto}. Since the relation $\kappa = 2\pi T$ follows from black hole thermodynamics, this result establishes a direct correspondence between the geometric description of particle motion and the quantum field theoretic interpretation of the chaos bound. 

This connection highlights a profound link between the instability of particle trajectories in black hole spacetimes and the universal constraints on chaotic growth. In this section, we explore possible violations of the chaos bound by examining the quantity $\lambda - \kappa$. A positive value of this difference signals a breakdown of the bound, whereas a value less than or equal to zero indicates that the bound remains satisfied.

\begin{figure}[h!]
    \centering
    \begin{subfigure}[b]{0.45\textwidth}
        \centering
        \includegraphics[width=\textwidth]{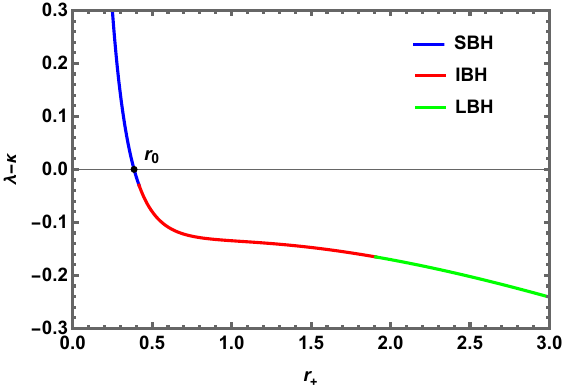}
        \caption{fixed $\epsilon=0.99$}
        \label{f6a}
    \end{subfigure}
    \hspace{0.8cm}
    \begin{subfigure}[b]{0.45\textwidth}
        \centering
        \includegraphics[width=\textwidth]{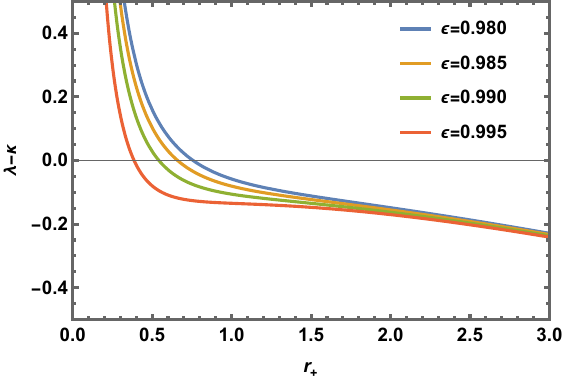}
        \caption{for different values of $\epsilon$}
        \label{f6b}
    \end{subfigure}
     \caption{Variation of $\lambda - \kappa$ with the horizon radius $r_+$, highlighting the regions where the chaos bound is violated for massless particles.}
    \label{f6}
\end{figure}
Figure \ref{f6a} illustrates the quantity $\lambda - \kappa$ as a function of the horizon radius for fixed value of $\epsilon$. The plot clearly indicates that the chaos bound is violated in the region where $r_+ < r_0$, with $r_0$ marked by the black dot in Fig.~\ref{f6a}. Furthermore, the violation occurs within the small black hole phase, depicted by the blue curve, which is thermodynamically stable. In contrast, for the remaining branches corresponding to non-small black hole phases, the bound is respected. 

Interestingly, it has been argued in \cite{Lei} that any violation of the chaos bound is restricted to thermodynamically stable regions of the black hole phase space. Our results are consistent with this observation: the interval over which $\lambda - \kappa > 0$ (beginning at $r_0 = 0.38632$) lies entirely within the stable phase and extends up to $r_+=0.41770$, beyond which the unstable intermediate black hole branch emerges. The stability of this region can be confirmed by verifying that the specific heat remains positive.

Figure~\ref{f6b} further displays the behaviour of $\lambda - \kappa$ as a function of the horizon radius $\tilde{r}_+$ for several values of the parameter $\epsilon$. From the plot, it is evident that decreasing $\epsilon$ shifts the point $r_0$, where $\lambda - \kappa$ first becomes positive (signalling the onset of chaos bound violation), toward smaller values of $r_+$. As a result, the interval of horizon radii over which the chaos bound is violated gradually shrinks with decreasing $\epsilon$.
\begin{figure}[h!]
        \includegraphics[width=0.5\textwidth]{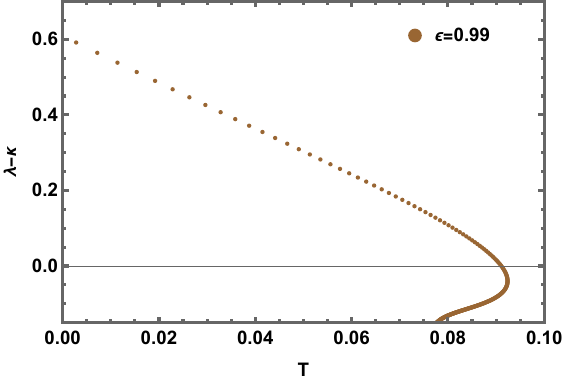}
        \caption{Variation of $\lambda - \kappa$ with the Hawking temperature $T$, highlighting the regions where the chaos bound is violated for massive particles}
        \label{f7}
\end{figure}
To examine the chaos bound in the case of massive particles, we plot the quantity $\lambda - \kappa$ as a function of the Hawking temperature (see figure \ref{f7}), since an explicit dependence of $\lambda$ on the horizon radius cannot be obtained, as discussed earlier. In this case as well, $\lambda - \kappa$ remains positive over a broad temperature range, indicating a violation of the chaos bound. However, unlike the massless case, it is not possible to clearly identify a threshold temperature at which this violation sets in.

The above discussions were for $\epsilon>\epsilon_c$ i.e., when a phase transition occurs. We may also examine the cases when no phase transitions occurs, when $\epsilon<\epsilon_c$.
\begin{figure}[h!]
    \centering
    \begin{subfigure}[b]{0.45\textwidth}
        \centering
        \includegraphics[width=\textwidth]{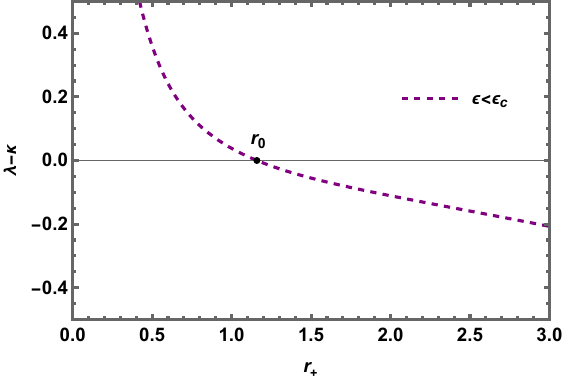}
        \caption{fixed $\epsilon<\epsilon_c$}
        \label{f8a}
    \end{subfigure}
    \hspace{0.8cm}
    \begin{subfigure}[b]{0.45\textwidth}
        \centering
        \includegraphics[width=\textwidth]{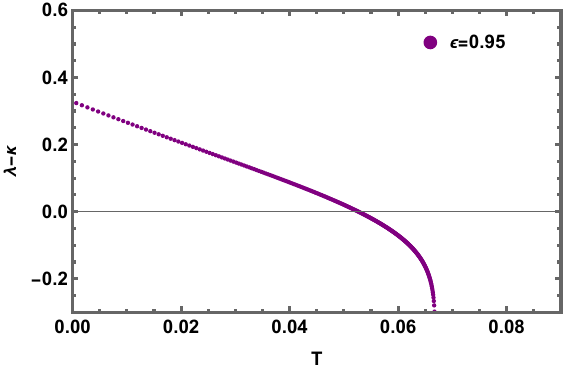}
        \caption{for massive particles, $\epsilon<\epsilon_c$}
        \label{f8b}
    \end{subfigure}
     \caption{Variation of $\lambda - \kappa$, highlighting the regions where the chaos bound is violated for massive particles.}
    \label{f8}
\end{figure}
Figures~\ref{f8a} and \ref{f8b} indicate that even in the absence of a phase transition, the chaos bound is violated for both massless and massive particles whenever the horizon radius drops below the critical value $r_0$.

It is useful to briefly reflect on the physical meaning of the chaos bound and its possible violation. Maldacena, Shenker, and Stanford showed that for systems with a large number of degrees of freedom admitting a semiclassical gravity dual, the growth of chaos - captured by out-of-time-ordered correlators - is universally bounded. In this picture, the Lyapunov exponent governs the rate of growth of perturbations, while the scrambling time $t_*$ characterizes how rapidly information spreads, with larger $\lambda$ implying faster scrambling.

However, as observed in several cases, including the present analysis, this bound can be violated. This suggests that, particularly in the thermodynamically stable small black hole phase, information scrambling may proceed more rapidly than predicted by semiclassical arguments. The persistence of this violation even in the absence of a phase transition ($\epsilon < \epsilon$) further indicates that it may originate from intrinsic features of the theory rather than being solely associated with critical phenomena.

\section{conclusion and discussion}\label{sec7}
Lyapunov exponent has been proven to be excellent probes of phase transitions in black holes arising in general relativity, however, its validity remains to be tested in modified and alternative theories of gravity. In this work we test the validity of the Lyapunov exponents as thermodynamic phase transition probes by applying to black holes in Horava Lifshitz gravity. 

In this work, we have investigated how Lyapunov exponents can be used to probe phase transitions in Horava Lifshitz AdS black holes. By evaluating the Lyapunov exponent for both massless and massive particles in circular orbits, we examined its dependence on the Hawking temperature. Our results show that when $\epsilon$ is above the critical value $\epsilon_c$, the Lyapunov exponent exhibits a multivalued profile as a function of temperature, with separate branches corresponding to the small, intermediate, and large black hole phases. As $\epsilon$ deceases below $\epsilon_c$, this multivalued behaviour disappears, closely resembling the corresponding trend in the free energy. This feature is observed for both types of particle motion, suggesting that the Lyapunov exponent effectively captures key aspects of the underlying thermodynamic phase structure.

We also analyzed the discontinuity in the Lyapunov exponent, $\Delta \lambda$, for both massless and massive probes, and showed that it effectively functions as an order parameter for the black hole phase transition. Our analytical results indicate that the corresponding critical exponent $\delta$ takes the value $1/2$, matching the mean-field prediction familiar from the van der Waals system. This conclusion is further supported by numerical evaluations of $\Delta \lambda / \lambda_c$ in the vicinity of the critical point, which display excellent agreement with the expected scaling behavior. These findings therefore highlight the robustness and universality of the critical exponent, consistently yielding $\delta = 1/2$.

We further investigate the chaos bound and its violation in Horava Lifshitz black hole background. We found that just like many exotic black hole systems, in the Horava Lifshitz background, the chaos bound is violated. The violation happens to occur in the thermodynamically stable and small black hole branch, whereas it s found to be preserved in the non-small black hole branches. It is observed that the chaos bound is always violated when the horizon radius is below a threshold. Further, we observe that decreasing the value of $\epsilon$ shifts the threshold point $r_0$ towards smaller values of horizon radius $r_+$. A possible reason for this could be that decreasing $\epsilon$ alters the background which changes the effective potential and the surface gravity such that $\lambda$ exceeds $\kappa$ at smaller radii. We also examine if the violation persists irrespective of whether there is a phase transition or not and found that the chaos bound is violated even when there is no phase transition.

Our results therefore indicate a strong correlation between the thermodynamic stability of the black hole and the breakdown of the chaos bound. This points to the possibility that stability characteristics play a crucial role in determining whether the bound is respected or violated. Clarifying this relationship could shed light on the intricate connections between black hole thermodynamics and quantum chaos. At the same time, the exact mechanism responsible for such violations, the nature of the underlying microscopic degrees of freedom, and their wider implications are still not fully understood. Resolving these issues is key to uncovering the fundamental principles that link black hole thermodynamics with chaotic dynamics - two seemingly distinct areas that are, in fact, deeply connected.


\begin{thebibliography}{99}

\bibitem{Phys}
S.~W.~Hawking,
``Gravitational radiation from colliding black holes,''
Phys. Rev. Lett. \textbf{26}, 1344-1346 (1971)
doi:10.1103/PhysRevLett.26.1344
\bibitem{bekens}
J.~Bekenstein,
``Bekenstein-Hawking entropy,''
Scholarpedia \textbf{3}, no.10, 7375 (2008)
doi:10.4249/scholarpedia.7375
\bibitem{Hawking}
S.~W.~Hawking,
``Black hole explosions,''
Nature \textbf{248}, 30-31 (1974)
doi:10.1038/248030a0
\bibitem{Hawking2}
S.~W.~Hawking,
``Particle Creation by Black Holes,''
Commun. Math. Phys. \textbf{43}, 199-220 (1975)
[erratum: Commun. Math. Phys. \textbf{46}, 206 (1976)]
doi:10.1007/BF02345020
\bibitem{Bardeen}
J.~M.~Bardeen, B.~Carter and S.~W.~Hawking,
``The Four laws of black hole mechanics,''
Commun. Math. Phys. \textbf{31}, 161-170 (1973)
doi:10.1007/BF01645742

\bibitem{Davies}
P.~C.~W.~Davies,
``Thermodynamics of Black Holes,''
Proc. Roy. Soc. Lond. A \textbf{353}, 499-521 (1977)
doi:10.1098/rspa.1977.0047
\bibitem{Hut}
P. Hut,``Charged black holes and phase transitions,”
Monthly Notices of the Royal
 Astronomical Society, vol. 180, pp. 379–389, 10 1977
 

\bibitem{Maldacena}
J.~M.~Maldacena,
``The Large N limit of superconformal field theories and supergravity,''
Adv. Theor. Math. Phys. \textbf{2}, 231-252 (1998)
doi:10.4310/ATMP.1998.v2.n2.a1
[arXiv:hep-th/9711200 [hep-th]].

\bibitem{Kubiz}
D.~Kubiznak and R.~B.~Mann,
``P-V criticality of charged AdS black holes,''
JHEP \textbf{07}, 033 (2012)
doi:10.1007/JHEP07(2012)033
[arXiv:1205.0559 [hep-th]].

\bibitem{Hawkpage}
S.~W.~Hawking and D.~N.~Page,
``Thermodynamics of Black Holes in anti-De Sitter Space,''
Commun. Math. Phys. \textbf{87}, 577 (1983)
doi:10.1007/BF01208266

\bibitem{Cai}
R.~G.~Cai, L.~M.~Cao, L.~Li and R.~Q.~Yang,
``P-V criticality in the extended phase space of Gauss-Bonnet black holes in AdS space,''
JHEP \textbf{09}, 005 (2013)
doi:10.1007/JHEP09(2013)005
[arXiv:1306.6233 [gr-qc]].

\bibitem{Kastor}
D.~Kastor, S.~Ray and J.~Traschen,
``Enthalpy and the Mechanics of AdS Black Holes,''
Class. Quant. Grav. \textbf{26}, 195011 (2009)
doi:10.1088/0264-9381/26/19/195011
[arXiv:0904.2765 [hep-th]].

\bibitem{Dolan}
B.~P.~Dolan,
``The cosmological constant and the black hole equation of state,''
Class. Quant. Grav. \textbf{28}, 125020 (2011)
doi:10.1088/0264-9381/28/12/125020
[arXiv:1008.5023 [gr-qc]].

\bibitem{Dolan2}
B.~P.~Dolan,
``Pressure and volume in the first law of black hole thermodynamics,''
Class. Quant. Grav. \textbf{28}, 235017 (2011)
doi:10.1088/0264-9381/28/23/235017
[arXiv:1106.6260 [gr-qc]].

\bibitem{Dolan3}
B.~P.~Dolan,
``Compressibility of rotating black holes,''
Phys. Rev. D \textbf{84}, 127503 (2011)
doi:10.1103/PhysRevD.84.127503
[arXiv:1109.0198 [gr-qc]].


\bibitem{Kubizna}
D.~Kubiznak, R.~B.~Mann and M.~Teo,
``Black hole chemistry: thermodynamics with Lambda,''
Class. Quant. Grav. \textbf{34}, no.6, 063001 (2017)
doi:10.1088/1361-6382/aa5c69
[arXiv:1608.06147 [hep-th]].

\bibitem{Xu}
W.~Xu, H.~Xu and L.~Zhao,
``Gauss-Bonnet coupling constant as a free thermodynamical variable and the associated criticality,''
Eur. Phys. J. C \textbf{74}, 2970 (2014)
doi:10.1140/epjc/s10052-014-2970-8
[arXiv:1311.3053 [gr-qc]].

\bibitem{Xu2}
W.~Xu and L.~Zhao,
``Critical phenomena of static charged AdS black holes in conformal gravity,''
Phys. Lett. B \textbf{736}, 214-220 (2014)
doi:10.1016/j.physletb.2014.07.019
[arXiv:1405.7665 [gr-qc]].

\bibitem{Zhang}
M.~Zhang, D.~C.~Zou and R.~H.~Yue,
``Reentrant phase transitions and triple points of topological AdS black holes in Born-Infeld-massive gravity,''
Adv. High Energy Phys. \textbf{2017}, 3819246 (2017)
doi:10.1155/2017/3819246
[arXiv:1707.04101 [hep-th]].


\bibitem{Ruppeiner:2012uc}
G.~Ruppeiner,
Thermodynamic curvature: pure fluids to black holes,
J. Phys. Conf. Ser. \textbf{410}, 012138 (2013)
doi:10.1088/1742-6596/410/1/012138
[arXiv:1210.2011 [gr-qc]].
\bibitem{Miao:2017cyt}
Y.~G.~Miao and Z.~M.~Xu,
Microscopic structures and thermal stability of black holes conformally coupled to scalar fields in five dimensions,
Nucl. Phys. B \textbf{942}, 205-220 (2019)
doi:10.1016/j.nuclphysb.2019.03.015
[arXiv:1711.01757 [hep-th]].
\bibitem{Guo:2019oad}
X.~Y.~Guo, H.~F.~Li, L.~C.~Zhang and R.~Zhao,
Microstructure and continuous phase transition of a Reissner-Nordstrom-AdS black hole,
Phys. Rev. D \textbf{100}, no.6, 064036 (2019)
doi:10.1103/PhysRevD.100.064036
[arXiv:1901.04703 [gr-qc]].
\bibitem{Wei:2019yvs}
S.~W.~Wei, Y.~X.~Liu and R.~B.~Mann,
Ruppeiner Geometry, Phase Transitions, and the Microstructure of Charged AdS Black Holes,
Phys. Rev. D \textbf{100}, no.12, 124033 (2019)
doi:10.1103/PhysRevD.100.124033
[arXiv:1909.03887 [gr-qc]].
\bibitem{Wang:2019cax}
P.~Wang, H.~Wu and H.~Yang,
Thermodynamic Geometry of AdS Black Holes and Black Holes in a Cavity,
Eur. Phys. J. C \textbf{80}, no.3, 216 (2020)
doi:10.1140/epjc/s10052-020-7776-2
[arXiv:1910.07874 [gr-qc]].
\bibitem{Yerra:2020oph}
P.~K.~Yerra and C.~Bhamidipati,
Ruppeiner Geometry, Phase Transitions and Microstructures of Black Holes in Massive Gravity,
Int. J. Mod. Phys. A \textbf{35}, no.22, 2050120 (2020)
doi:10.1142/S0217751X20501201
[arXiv:2006.07775 [hep-th]].
\bibitem{Yerra:2021hnh}
P.~K.~Yerra and C.~Bhamidipati,
Novel relations in massive gravity at Hawking-Page transition,
Phys. Rev. D \textbf{104}, no.10, 104049 (2021)
doi:10.1103/PhysRevD.104.104049
[arXiv:2107.04504 [gr-qc]].


\bibitem{Wu:2022whe}
D.~Wu,
Topological classes of rotating black holes,
Phys. Rev. D \textbf{107}, no.2, 024024 (2023)
doi:10.1103/PhysRevD.107.024024
[arXiv:2211.15151 [gr-qc]].
\bibitem{Liu:2022aqt}
C.~Liu and J.~Wang,
Topological natures of the Gauss-Bonnet black hole in AdS space,
Phys. Rev. D \textbf{107}, no.6, 064023 (2023)
doi:10.1103/PhysRevD.107.064023
[arXiv:2211.05524 [gr-qc]].
\bibitem{Fan:2022bsq}
Z.~Y.~Fan,
Topological interpretation for phase transitions of black holes,
Phys. Rev. D \textbf{107}, no.4, 044026 (2023)
doi:10.1103/PhysRevD.107.044026
[arXiv:2211.12957 [gr-qc]].

\bibitem{Gogoi:2023xzy}
N.~J.~Gogoi and P.~Phukon,
Thermodynamic topology of 4D dyonic AdS black holes in different ensembles,
Phys. Rev. D \textbf{108}, no.6, 066016 (2023)
doi:10.1103/PhysRevD.108.066016
[arXiv:2304.05695 [hep-th]].



\bibitem{Ali:2023zww}
M.~S.~Ali, H.~El Moumni, J.~Khalloufi and K.~Masmar,
Topology of Born-Infeld-AdS Black Hole Phase Transition,
[arXiv:2306.11212 [hep-th]].

\bibitem{Saleem:2023oue}
M.~A.~Saleem and A.~Taani,
The chaotic behavior of black holes: Investigating a topological retraction in anti-de Sitter spaces,
New Astron. \textbf{107}, 102149 (2024)
doi:10.1016/j.newast.2023.102149
\bibitem{Shahzad:2023cis}
M.~U.~Shahzad, A.~Mehmood, S.~Sharif and A.~\"Ovg\"un,
Criticality and topological classes of neutral Gauss\textendash{}Bonnet AdS black holes in 5D,
Annals Phys. \textbf{458}, no.3, 169486 (2023)
doi:10.1016/j.aop.2023.169486
\bibitem{Chen:2023elp}
Z.~Q.~Chen and S.~W.~Wei,
Thermodynamics, Ruppeiner geometry, and topology of Born-Infeld black hole in asymptotic flat spacetime,
Nucl. Phys. B \textbf{996}, 116369 (2023)
doi:10.1016/j.nuclphysb.2023.116369
\bibitem{Bai:2022klw}
N.~C.~Bai, L.~Li and J.~Tao,
Topology of black hole thermodynamics in Lovelock gravity,
Phys. Rev. D \textbf{107}, no.6, 064015 (2023)
doi:10.1103/PhysRevD.107.064015
[arXiv:2208.10177 [gr-qc]].
\bibitem{Yerra:2022alz}
P.~K.~Yerra and C.~Bhamidipati,
Topology of black hole thermodynamics in Gauss-Bonnet gravity,
Phys. Rev. D \textbf{105}, no.10, 104053 (2022)
doi:10.1103/PhysRevD.105.104053
[arXiv:2202.10288 [gr-qc]].
\bibitem{Hazarika:2023iwp}
B.~Hazarika and P.~Phukon,
Thermodynamic Topology of $D=4,5$ Horava Lifshitz Black Hole in Two Ensembles,
[arXiv:2312.06324 [hep-th]].


\bibitem{Liu:2014gvf}
Y.~Liu, D.~C.~Zou and B.~Wang,
Signature of the Van der Waals like small-large charged AdS black hole phase transition in quasinormal modes,
JHEP \textbf{09}, 179 (2014)
doi:10.1007/JHEP09(2014)179
[arXiv:1405.2644 [hep-th]].
\bibitem{Zou:2017juz}
D.~C.~Zou, Y.~Liu and R.~H.~Yue,
Behavior of quasinormal modes and Van der Waals-like phase transition of charged AdS black holes in massive gravity,
Eur. Phys. J. C \textbf{77}, no.6, 365 (2017)
doi:10.1140/epjc/s10052-017-4937-z
[arXiv:1702.08118 [gr-qc]].
\bibitem{Zhang:2020khz}
M.~Zhang, C.~M.~Zhang, D.~C.~Zou and R.~H.~Yue,
Phase transition and Quasinormal modes for Charged black holes in 4D Einstein-Gauss-Bonnet gravity,
Chin. Phys. C \textbf{45}, no.4, 045105 (2021)
doi:10.1088/1674-1137/abe19a
[arXiv:2009.03096 [hep-th]].
\bibitem{Mahapatra:2016dae}
S.~Mahapatra,
Thermodynamics, Phase Transition and Quasinormal modes with Weyl corrections,
JHEP \textbf{04}, 142 (2016)
doi:10.1007/JHEP04(2016)142
[arXiv:1602.03007 [hep-th]].
\bibitem{Chabab:2016cem}
M.~Chabab, H.~El Moumni, S.~Iraoui and K.~Masmar,
Behavior of quasinormal modes and high dimension RN\textendash{}AdS black hole phase transition,
Eur. Phys. J. C \textbf{76}, no.12, 676 (2016)
doi:10.1140/epjc/s10052-016-4518-6
[arXiv:1606.08524 [hep-th]].


\bibitem{Wei:2017mwc}
S.~W.~Wei and Y.~X.~Liu,
Photon orbits and thermodynamic phase transition of $d$-dimensional charged AdS black holes,
Phys. Rev. D \textbf{97}, no.10, 104027 (2018)
doi:10.1103/PhysRevD.97.104027
[arXiv:1711.01522 [gr-qc]].
\bibitem{Wei:2018aqm}
S.~W.~Wei, Y.~X.~Liu and Y.~Q.~Wang,
Probing the relationship between the null geodesics and thermodynamic phase transition for rotating Kerr-AdS black holes,
Phys. Rev. D \textbf{99}, no.4, 044013 (2019)
doi:10.1103/PhysRevD.99.044013
[arXiv:1807.03455 [gr-qc]].
\bibitem{Zhang:2019tzi}
M.~Zhang, S.~Z.~Han, J.~Jiang and W.~B.~Liu,
Phys. Rev. D \textbf{99}, no.6, 065016 (2019)
doi:10.1103/PhysRevD.99.065016
[arXiv:1903.08293 [hep-th]].



\bibitem{Zhang:2019glo}
M.~Zhang and M.~Guo,
Can shadows reflect phase structures of black holes?,
Eur. Phys. J. C \textbf{80}, no.8, 790 (2020)
doi:10.1140/epjc/s10052-020-8389-5
[arXiv:1909.07033 [gr-qc]].
\bibitem{Belhaj:2020nqy}
A.~Belhaj, L.~Chakhchi, H.~El Moumni, J.~Khalloufi and K.~Masmar,
Thermal Image and Phase Transitions of Charged AdS Black Holes using Shadow Analysis,
Int. J. Mod. Phys. A \textbf{35}, no.27, 2050170 (2020)
doi:10.1142/S0217751X20501705
[arXiv:2005.05893 [gr-qc]].

\bibitem{lyp}
A.M Lyapunov,
The general problem of the stability of motion
Int. J. Control 55, 531 (1992).
doi:https://doi.org/10.1080/00207179208934253
\bibitem{lyp2}
M. Sandri,
Numerical calculation of lyapunov exponents,
Math. J. 6, 78 (1996).


\bibitem{syk}
N.~Sorokhaibam,
``Phase transition and chaos in charged SYK model,''
JHEP \textbf{07}, 055 (2020)
doi:10.1007/JHEP07(2020)055
[arXiv:1912.04326 [hep-th]].
\bibitem{syk2}
A.~Davis and Y.~Wang,
``Quantum chaos and phase transition in the Yukawa\textendash{}Sachdev-Ye-Kitaev model,''
Phys. Rev. B \textbf{107}, no.20, 205122 (2023)
doi:10.1103/PhysRevB.107.205122
[arXiv:2212.03265 [cond-mat.str-el]].
\bibitem{Dicke}
C.~Emary and T.~Brandes,
``Chaos and the quantum phase transition in the Dicke model,''
Phys. Rev. E \textbf{67}, 066203 (2003)
doi:10.1103/PhysRevE.67.066203
[arXiv:cond-mat/0301273 [cond-mat]].
\bibitem{coscll}
G.~Miritello, A.~Pluchino and A.~Rapisarda,
``Phase Transitions and Chaos in Long-Range Models of Coupled Oscillators,''
EPL \textbf{85}, no.1, 10007 (2009)
doi:10.1209/0295-5075/85/10007
[arXiv:0807.1870 [cond-mat.stat-mech]].
\bibitem{finite}
W.~D.~Heiss and A.~L.~Sannino,
``Transitional regions of finite Fermi systems and quantum chaos,''
Phys. Rev. A \textbf{43}, 4159-4166 (1991)
doi:10.1103/PhysRevA.43.4159


\bibitem{static}
Y.~Sota, S.~Suzuki and K.~i.~Maeda,
``Chaos in static axisymmetric space-times. 1: Vacuum case,''
Class. Quant. Grav. \textbf{13}, 1241-1260 (1996)
doi:10.1088/0264-9381/13/5/034
[arXiv:gr-qc/9505036 [gr-qc]].
\bibitem{static2}
Y.~Sota, S.~Suzuki and K.~i.~Maeda,
``Chaos in static axisymmetric space-times. 2. Nonvacuum case,''
[arXiv:gr-qc/9610065 [gr-qc]].
\bibitem{kerr}
N.~Kan and B.~Gwak,
``Bound on the Lyapunov exponent in Kerr-Newman black holes via a charged particle,''
Phys. Rev. D \textbf{105}, no.2, 026006 (2022)
doi:10.1103/PhysRevD.105.026006
[arXiv:2109.07341 [gr-qc]].
\bibitem{kerr2}
B.~Gwak, N.~Kan, B.~H.~Lee and H.~Lee,
``Violation of bound on chaos for charged probe in Kerr-Newman-AdS black hole,''
JHEP \textbf{09}, 026 (2022)
doi:10.1007/JHEP09(2022)026
[arXiv:2203.07298 [gr-qc]].
\bibitem{multi}
W.~Hanan and E.~Radu,
``Chaotic motion in multi-black hole spacetimes and holographic screens,''
Mod. Phys. Lett. A \textbf{22}, 399-406 (2007)
doi:10.1142/S0217732307022815
[arXiv:gr-qc/0610119 [gr-qc]].
\bibitem{qg}
F.~Lu, J.~Tao and P.~Wang,
``Minimal Length Effects on Chaotic Motion of Particles around Black Hole Horizon,''
JCAP \textbf{12}, 036 (2018)
doi:10.1088/1475-7516/2018/12/036
[arXiv:1811.02140 [gr-qc]].
\bibitem{qg2}
X.~Guo, K.~Liang, B.~Mu, P.~Wang and M.~Yang,
``Chaotic Motion around a Black Hole under Minimal Length Effects,''
Eur. Phys. J. C \textbf{80}, no.8, 745 (2020)
doi:10.1140/epjc/s10052-020-8335-6
[arXiv:2002.05894 [gr-qc]].

\bibitem{mss}
J.~Maldacena, S.~H.~Shenker and D.~Stanford,
``A bound on chaos,''
JHEP \textbf{08}, 106 (2016)
doi:10.1007/JHEP08(2016)106
[arXiv:1503.01409 [hep-th]].
\bibitem{hori}
K.~Hashimoto and N.~Tanahashi,
``Universality in Chaos of Particle Motion near Black Hole Horizon,''
Phys. Rev. D \textbf{95}, no.2, 024007 (2017)
doi:10.1103/PhysRevD.95.024007
[arXiv:1610.06070 [hep-th]].
\bibitem{hori2}
S.~Dalui, B.~R.~Majhi and P.~Mishra,
``Presence of horizon makes particle motion chaotic,''
Phys. Lett. B \textbf{788}, 486-493 (2019)
doi:10.1016/j.physletb.2018.11.050
[arXiv:1803.06527 [gr-qc]].

\bibitem{vio}
Q.~Q.~Zhao, Y.~Z.~Li and H.~Lu,
``Static Equilibria of Charged Particles Around Charged Black Holes: Chaos Bound and Its Violations,''
Phys. Rev. D \textbf{98}, no.12, 124001 (2018)
doi:10.1103/PhysRevD.98.124001
[arXiv:1809.04616 [gr-qc]].
\bibitem{vio2}
X.~Guo, K.~Liang, B.~Mu, P.~Wang and M.~Yang,
``Minimal Length Effects on Motion of a Particle in Rindler Space,''
Chin. Phys. C \textbf{45}, no.2, 023115 (2021)
doi:10.1088/1674-1137/abcf20
[arXiv:2007.07744 [gr-qc]].
\bibitem{vio3}
B.~Gwak, N.~Kan, B.~H.~Lee and H.~Lee,
``Violation of bound on chaos for charged probe in Kerr-Newman-AdS black hole,''
JHEP \textbf{09}, 026 (2022)
doi:10.1007/JHEP09(2022)026
[arXiv:2203.07298 [gr-qc]].
\bibitem{vio4}
J.~Park and B.~Gwak,
``Bound on Lyapunov exponent in Kerr-Newman-de Sitter black holes by a charged particle,''
JHEP \textbf{04}, 023 (2024)
doi:10.1007/JHEP04(2024)023
[arXiv:2312.13075 [gr-qc]].
\bibitem{first}
X.~Guo, Y.~Lu, B.~Mu and P.~Wang,
``Probing phase structure of black holes with Lyapunov exponents,''
JHEP \textbf{08}, 153 (2022)
doi:10.1007/JHEP08(2022)153
[arXiv:2205.02122 [gr-qc]].
\bibitem{le}
S.~Yang, J.~Tao, B.~Mu and A.~He,
``Lyapunov exponents and phase transitions of Born-Infeld AdS black holes,''
JCAP \textbf{07}, 045 (2023)
doi:10.1088/1475-7516/2023/07/045
[arXiv:2304.01877 [gr-qc]].
\bibitem{le2}
X.~Lyu, J.~Tao and P.~Wang,
``Probing the thermodynamics of charged Gauss Bonnet AdS black holes with the Lyapunov exponent,''
Eur. Phys. J. C \textbf{84}, no.9, 974 (2024)
doi:10.1140/epjc/s10052-024-13354-9
[arXiv:2312.11912 [gr-qc]].
\bibitem{le3}
A.~N.~Kumara, S.~Punacha and M.~S.~Ali,
``Lyapunov exponents and phase structure of Lifshitz and hyperscaling violating black holes,''
JCAP \textbf{07}, 061 (2024)
doi:10.1088/1475-7516/2024/07/061
[arXiv:2401.05181 [gr-qc]].
\bibitem{le4}
Y.~Z.~Du, H.~F.~Li, Y.~B.~Ma and Q.~Gu,
``Phase structure and optical properties of the de Sitter Spacetime with KR field based on the Lyapunov exponent,''
Eur. Phys. J. C \textbf{85}, no.1, 78 (2025)
doi:10.1140/epjc/s10052-025-13809-7
[arXiv:2403.20083 [hep-th]].
\bibitem{le5}
N.~J.~Gogoi, S.~Acharjee and P.~Phukon,
``Lyapunov exponents and phase transition of Hayward AdS black hole,''
Eur. Phys. J. C \textbf{84}, no.11, 1144 (2024)
doi:10.1140/epjc/s10052-024-13520-z
[arXiv:2404.03947 [hep-th]].
\bibitem{le6}
B.~Shukla, P.~P.~Das, D.~Dudal and S.~Mahapatra,
``Interplay between the Lyapunov exponents and phase transitions of charged AdS black holes,''
Phys. Rev. D \textbf{110}, no.2, 024068 (2024)
doi:10.1103/PhysRevD.110.024068
[arXiv:2404.02095 [hep-th]].
\bibitem{le7}
D.~Chen, C.~Yang and Y.~Liu,
``Lyapunov exponents as probes for a phase transition of a Kerr-AdS black hole,''
Phys. Lett. B \textbf{865}, 139463 (2025)
doi:10.1016/j.physletb.2025.139463
[arXiv:2501.16999 [hep-th]].
\bibitem{le8}
K.~R., D.~D., K.~M.~Ajith, K.~Hegde, S.~Punacha and A.~N.~Kumara,
``Euclidean Thermodynamics and Lyapunov Exponents of Einstein-Power-Yang-Mills AdS Black Holes,''
[arXiv:2504.12890 [gr-qc]].
\bibitem{Awal}
M.~B.~Awal and P.~Phukon,
``Probing Thermodynamic Phase Transitions of 4D R-Charged Black Holes via Lyapunov Exponent,''
[arXiv:2505.20800 [gr-qc]].
\bibitem{le9}
C.~Yang, C.~Gao, D.~Chen and X.~Zeng,
``Lyapunov exponents, phase transition and chaos bound in Kerr-Newman AdS spacetime,''
[arXiv:2506.21882 [hep-th]].
\bibitem{le10}
X.~Guo, R.~Yang, Y.~Liang and J.~Tao,
``Lyapunov exponents and phase transition of charged Ads black hole in quintessence,''
[arXiv:2508.03519 [gr-qc]].
\bibitem{mba}
G.~Bezboruah, M.~B.~Awal and P.~Phukon,
``Lyapunov exponents, phase transitions, and Chaos bound of ModMax AdS black holes,''
Eur. Phys. J. C \textbf{85} (2025) no.10, 1169
doi:10.1140/epjc/s10052-025-14920-5
[arXiv:2508.07832 [gr-qc]].


\bibitem{Horava:2009uw}
P.~Horava,
``Quantum Gravity at a Lifshitz Point,''
Phys. Rev. D \textbf{79} (2009), 084008
doi:10.1103/PhysRevD.79.084008
[arXiv:0901.3775 [hep-th]].

\bibitem{ht1}
R.~G.~Cai, L.~M.~Cao and N.~Ohta,
``Topological Black Holes in Horava-Lifshitz Gravity,''
Phys. Rev. D \textbf{80} (2009), 024003
doi:10.1103/PhysRevD.80.024003
[arXiv:0904.3670 [hep-th]].
\bibitem{ht2}
R.~G.~Cai, L.~M.~Cao and N.~Ohta,
``Thermodynamics of Black Holes in Horava-Lifshitz Gravity,''
Phys. Lett. B \textbf{679} (2009), 504-509
doi:10.1016/j.physletb.2009.07.075
[arXiv:0905.0751 [hep-th]].
\bibitem{ht3}
Q.~J.~Cao, Y.~X.~Chen and K.~N.~Shao,
``Black hole phase transitions in Ho{\v{r}}ava-Lifshitz gravity,''
Phys. Rev. D \textbf{83} (2011), 064015
doi:10.1103/PhysRevD.83.064015
[arXiv:1010.5044 [hep-th]].
\bibitem{ht4}
B.~R.~Majhi and D.~Roychowdhury,
``Phase transition and scaling behavior of topological charged black holes in Horava-Lifshitz gravity,''
Class. Quant. Grav. \textbf{29} (2012), 245012
doi:10.1088/0264-9381/29/24/245012
[arXiv:1205.0146 [gr-qc]].
\bibitem{Li}
T.~J.~Li, Y.~H.~Qi, Y.~L.~Wu and Y.~L.~Zhang,
``Topological charged black holes in generalized Ho{\v{r}}ava-Lifshitz gravity,''
Phys. Rev. D \textbf{90} (2014) no.12, 124070
doi:10.1103/PhysRevD.90.124070
[arXiv:1405.4457 [hep-th]].



\bibitem{ce}
R.~Banerjee and D.~Roychowdhury,
``Critical behavior of Born Infeld AdS black holes in higher dimensions,''
Phys. Rev. D \textbf{85} (2012), 104043
doi:10.1103/PhysRevD.85.104043
[arXiv:1203.0118 [gr-qc]].






\bibitem{Hashimoto}
K.~Hashimoto and N.~Tanahashi,
``Universality in Chaos of Particle Motion near Black Hole Horizon,''
Phys. Rev. D \textbf{95} (2017) no.2, 024007
doi:10.1103/PhysRevD.95.024007
[arXiv:1610.06070 [hep-th]].

\bibitem{Lei}
Y.~Q.~Lei, X.~H.~Ge and S.~Dalui,
``Thermodynamic stability versus chaos bound violation in D-dimensional RN black holes: Angular momentum effects and phase transitions,''
Phys. Lett. B \textbf{856}, 138929 (2024)
doi:10.1016/j.physletb.2024.138929
[arXiv:2404.18193 [hep-th]].








\end{thebibliography}
\end{document}